\documentclass[letter]{aa}
\usepackage{natbib}
\usepackage{graphicx}
\usepackage{txfonts}

\usepackage{color}

\newcommand{\Gaia}{\textit{Gaia}\ }

\begin{document}

\title{VLBI-\Gaia offsets favor parsec-scale jet direction in \\ Active Galactic Nuclei}
\titlerunning{VLBI-\Gaia offsets favor AGN jet direction} 

\author{Y.~Y.~Kovalev\inst{1,2}
\and
L.~Petrov\inst{3}
\and
A.~V.~Plavin\inst{2,4}
}
\authorrunning{Kovalev, Petrov, \& Plavin}
\institute{
Max-Planck-Institut f\"ur Radioastronomie, Auf dem H\"ugel 69, 53121 Bonn, Germany
\and
Astro Space Center of Lebedev Physical Institute, Profsoyuznaya 86/32, 117997 Moscow, Russia; \email{yyk@asc.rssi.ru}
\and
Astrogeo Center, 7312 Sportsman Dr., Falls Church, VA 22043, USA
\and
Moscow Institute of Physics and Technology, Institutsky per. 9, Dolgoprudny 141700, Russia
}

\date{Received 8~November~2016; accepted 8~December~2016}

\abstract
{
The data release 1 (DR1) of milliarcsecond-scale accurate optical positions of stars and galaxies was recently published by the space mission \textit{Gaia}.
}
{
We study the offsets of highly accurate absolute radio (very long baseline interferometry, VLBI) and optical positions of active galactic nuclei (AGN) to see whether or not a signature of wavelength-dependent parsec-scale structure can be seen.
}
{
We analyzed VLBI and \Gaia positions and determined the direction of jets in 2957 AGNs from their VLBI images.
}
{
We find that there is a statistically significant excess of sources with VLBI-to-\Gaia position offset in directions along and opposite to the jet.
Offsets along the jet vary from 0 to tens of mas. Offsets in the opposite direction do not exceed 3~mas.
}
{
The presense of strong, extended parsec-scale optical jet structures in many AGNs is required to explain all observed VLBI-\Gaia offsets along the jet direction.
The offsets in the opposite direction shorter than 1~mas can be explained either by a non-point-like VLBI jet structure or a ``core-shift'' effect due to synchrotron opacity.
}


\maketitle

\section{Introduction}
\label{s:intro}

\Gaia Data Release~1 \citep[DR1,][]{Gaia_DR1_astrometry} provides a catalogue of highly accurate optical positions for many objects, including active galactic nuclei (AGNs) with milliarcsecond uncertainties. So far, only VLBI has been able to provide that level of accuracy. A comparison of the \Gaia quasar auxiliary solution with the International Celestial Reference Frame~2 (ICRF2) catalog \citep{ICRF2} demonstrated a good agreement in radio and optic positions, but singled out a fraction of 6\% as outliers \citep{Gaia_DR1_frame}. 
\citet{PK16} extended this comparison to the secondary \Gaia DR1 catalogue of 1.14 billion objects that have median position uncertainty 2.3~mas and modern VLBI absolute astrometry catalogue RFC~2016c\footnote{\url{http://astrogeo.org/vlbi/solutions/rfc_2016c/}} that, to date, is the most complete. 
They found 6055 firm matches with AGNs.
Both ICRF2 and RFC~2016c catalogues have comparable accuracy, but the latter utilized all VLBI observations used for the ICRF2 and those that became available from January~2008 through September 2016. This increased the total number of VLBI sources by more than a factor of three with respect to the ICRF2. \citet{PK16} revealed a population of approximately 400 objects with significant radio/optical offsets after alignment of two catalogs that cannot be explained by the random noise in the data. However, that study could not provide information on the cause of these offsets.

This motivated us to consider additional available information about AGN 
structure at milliarcsec scale that can shed light on the cause of the offsets 
between the radio absolute reference points and the \Gaia centroids in the optical band.
The majority of radio-loud AGNs exhibit a typical core-jet morphology, thus presenting a strong
asymmetry in their structure. For many of them, the jet is resolved and strong enough for us to determine its direction from hybrid mapping results. Let us define the offset vector of 
the \Gaia position with respect to the VLBI position $\vec{VG}$ and the unit vector, 
defining the direction of a jet from the jet base downstream as $\vec{j}$
(Fig.~\ref{f:diagram}).
The angular difference between these directions is denoted as 
$\Delta\mathrm{P.A.^{VG}_{jet}}$. In the following, we analyze the distribution 
of $\Delta\mathrm{P.A.^{VG}_{jet}}$ for radio/optical AGN matches and analyze its 
implications. 

\section{Observational data and basic analysis}
\label{s:data-anal}

\begin{figure}[b]
\centering
\includegraphics[width=0.40\textwidth,trim=0cm 0.7cm 0cm 0.3cm,clip]{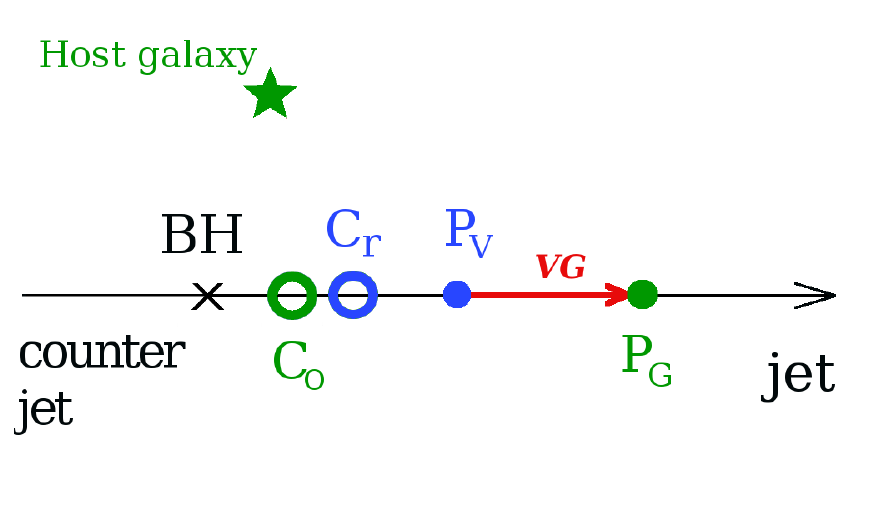}
\caption{
A simplified diagram of an AGN at milliarcsecond scales in the plane of the sky.
The `BH' cross marks a position of the supermassive black hole as well as the accretion disk. The arrow represents the jet ($\vec{j}$ vector) while the counter-jet goes in the opposite direction.
The apparent base of the jet in radio is shown by `$\mathrm{C_r}$', in the optical band it is expected to be closer to the central engine `BH' \citep{Kovalev_cs_2008} and is shown as `$\mathrm{C_o}$'. The absolute radio VLBI `$\mathrm{P_V}$' and optical \Gaia `$\mathrm{P_G}$' reference points are shown by blue and green dots, respectively. The red offset vector $\vec{VG}$ connects these points. The host galaxy can be relatively bright in optical band and shift the optical centroid in any direction, the galaxy center is marked by a star.
\label{f:diagram}
}
\end{figure}

Let us consider a simplified AGN diagram in Fig.~\ref{f:diagram}. It was shown that 
the apparent base of the jet in radio band `$\mathrm{C_r}$' typically associated with the brightest and most compact region in AGN jets at parsec scales changes its position with frequency due to the synchrotron self-absorption 
\citep{MS84,L98}. Observations demonstrated that the core-shift is typically 
at a sub-mas level at centimeter wavelengths \cite[e.g.,][]{Sokol_cs2011,pushkarev_etal12,fromm_etal13,kutkin_etal14}. 
\citet{Kovalev_cs_2008} predicted that the apparent jet base in the optical band `$\mathrm{C_o}$' will be shifted at 0.1~mas level with respect to the jet base `$\mathrm{C_r}$' at 8~GHz in the direction to `BH'. 
However, if the core-shift depends on frequency as $\nu^{-1}$, it has 
no contribution to group delay that is used for absolute VLBI astrometry  
\citep{Porcas_cs2009} and thus, does not affect the absolute VLBI positions. In that particular case `$\mathrm{C_r}$' and `$\mathrm{C_o}$' coincide.

\begin{figure*}[t]
\centering
\includegraphics[width=0.24\textwidth,angle=0,trim=0.35cm 0.2cm 0.2cm 0.3cm,clip]{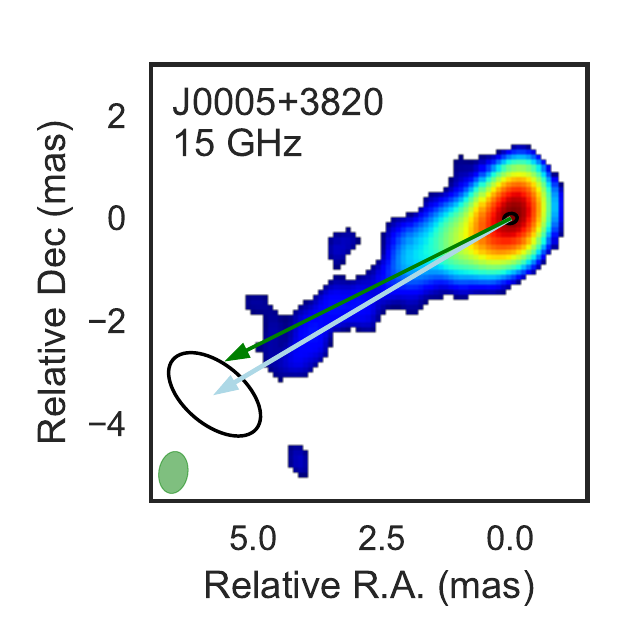}
\includegraphics[width=0.24\textwidth,angle=0,trim=0.35cm 0.2cm 0.2cm 0.3cm,clip]{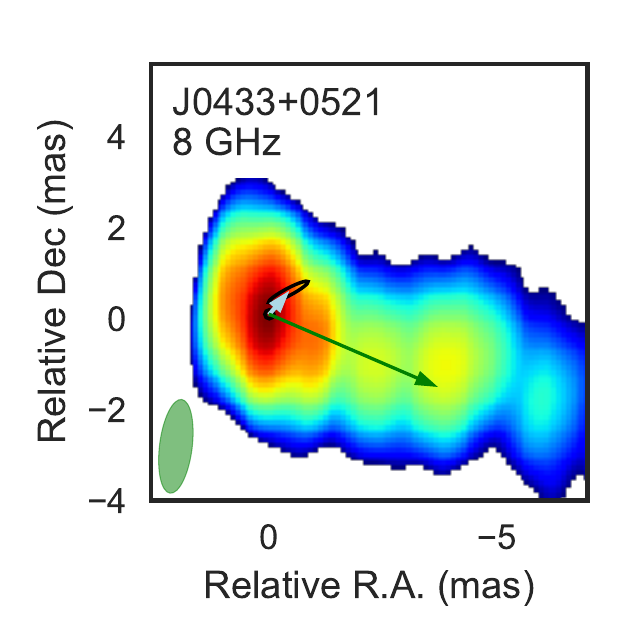}
\includegraphics[width=0.24\textwidth,angle=0,trim=0.35cm 0.2cm 0.2cm 0.3cm,clip]{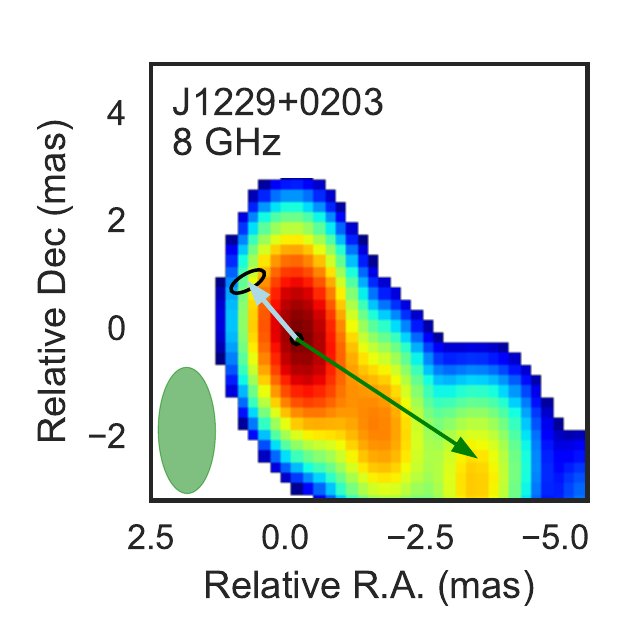}
\includegraphics[width=0.24\textwidth,angle=0,trim=0.35cm 0.2cm 0.2cm 0.3cm,clip]{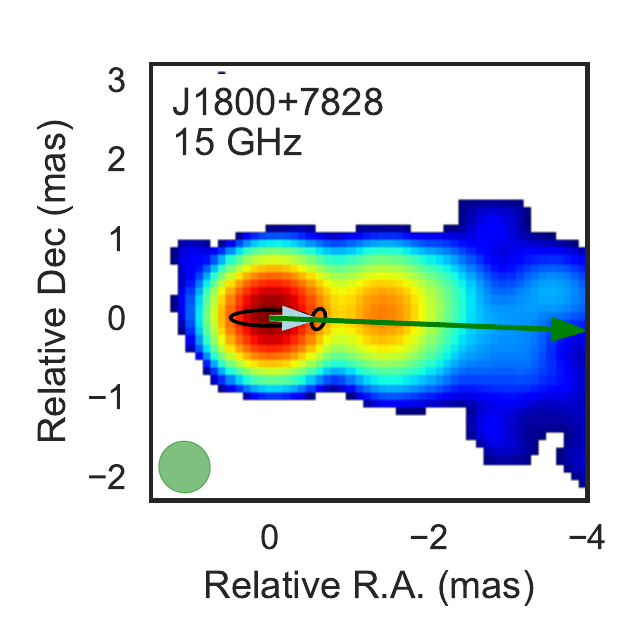}
\caption{
VLBI images of brightness distributions at 8 and 15~GHz are shown together with estimated median direction of their jets (green arrow, the $\vec{j}$ vector) for the four targets out of the total of 2957 used in the analysis.
Low-to-high surface brightness is shown by color from dark blue to dark red.
The VLBI beam is shown by the light green ellipse at the half power level in the bottom left corner.
The relative position of the peak intensity pixel on the map is formally chosen as the VLBI astrometry reference point of the target to illustrate the effect.
The vector $\vec{VG}$ is shown in light blue relative to this location from VLBI-to-\Gaia position and black ellipses represent $1\sigma$ errors.
$\Delta\mathrm{P.A.^{VG}_{jet}}$ is measured to be $4^\circ$, $73^\circ$, $164^\circ$, and $1^\circ$ for J0005+3820, J0433+0521 (3C\,120), J1229+0203 (3C\,273), and J1800+7828 (S5\,1803+784), respectively.
\label{f:maps}
}
\end{figure*}

The presence of the asymmetric radio structure causes an additional term in
group delay \citep{Charlot90}. This term is ignored in both ICRF2 and RFC~2016c
data analyses. Neglecting this term results in a shift of the VLBI reference
point `$\mathrm{P_V}$' with respect to the jet base, predominantly
in the direction along the jet (Fig.~\ref{f:diagram}). \citet{C02,S02} 
investigated this effect in detail and found that an unaccounted-for  structure term
causes source position jitter with the rms that, on average, does not exceed 
0.11~mas, but for extreme cases can cause peak-to-peak variations of up to 2~mas.
It should be noted that AGNs are very active, sometimes flaring objects with 
physical conditions changing dramatically in regions close to the nucleus. Thus,
the distance between the apparent jet base `$\mathrm{C}_\mathrm{r}$' and the reference point `$\mathrm{P_V}$' may change.
Analysis of source position time series \citep{r:fes00} found that variations 
in source positions caused by unaccounted-for changes in source structure rarely 
exceed 1~mas. 
Strong scattering of radio emission could affect positions \citep[e.g.,][]{pushkarev_etal13} but for most targets can be neglected \citep{puskov15}.
A counter-jet is observed in a small fraction of AGNs \citep[e.g.,][]{LPA16} in addition to the main jet, although usually it is weak due to de-boosting.

Similarly, milliarcsecond-scale structure of AGNs affects the position of the centroid in the optical band. Some active galaxies are known to have extended and bright jets \citep[e.g.,][]{Prieto16,2000ApJ...542..731F} at scales of hundreds of parsecs. 
As a result, the optical centroid position is shifted along the jet from `$\mathrm{C}_\mathrm{o}$' to `$\mathrm{P_G}$' (Fig.~\ref{f:diagram}) by values that could exceed the 1-mas scale.
It is important to note that VLBI positions are determined on the basis of VLBI visibility measurements, not sensitive to extended structures, while \Gaia detects total power. 
For this reason, the extended optical emission affects the source position 
differently than extended radio emission.
Additionally, an accretion disk may have the optical emission centered at the super-massive black hole `BH', which may be shifted at a level of a fraction of a milliarcsecond with respect to the apparent jet base `$\mathrm{C}_\mathrm{o}$' in the direction opposite to the jet.
The optical `host galaxy' center of mass might be shifted from `BH' in any direction on the milliarcsecond scale.


\cite{PK16} have associated the \Gaia DR1 secondary dataset and the VLBI RFC~2016c catalogs.
For the following analysis, we selected 6054 matches with the probability of false association $\mathrm{PFA}< 2\cdot10^{-4}$. 
PFA was calculated using \Gaia source density averaged within a cell of a regular $0\fdg25\times0\fdg25$ grid \citep{PK16}.
The 50th percentile of \textit{Gaia}~--~RFC offset lengths is 2.2~mas and the 99th is 76~mas.

The position angle $\mathrm{P.A._{jet}}$ of the parsec-scale radio jet, vector $\vec{j}$, is determined using VLBI images from the Astrogeo VLBI FITS image database\footnote{\url{http://astrogeo.org/vlbi_images/}}.
The images that we used come mostly from the analysis of the VLBA Calibrator Survey \citep[VCS;][]{vcs1,vcs2,vcs3,vcs4,vcs5,vcs6} and regular geodesy VLBI program \citep{RDV2009,puskov12,Piner12} at 2 and 8~GHz. Additionally, we made some use of images from the VLBI Imaging and Polarimetry Survey at 5~GHz \citep{Hel07,Petrov_VIPS}, the VCS releases 7,~8, and~9 \citep{VCS9} at 7.4~GHz; VLBI observing programs for \textit{Fermi}-AGN associations \citep[e.g.,][]{schinzel15} at 8~GHz; the 15~GHz Monitoring Of Jets in Active Galactic Nuclei with VLBA Experiments program \citep[\mbox{MOJAVE},][]{MOJAVE_V}, 24 and 43~GHz images from the K/Q survey \citep{Charlot10} and the VLBA-BU Blazar Monitoring Program \citep{BU16}.
%
%
The jet direction $\vec{j}$ is determined from the inner direction of the jet ridge line calculated directly from the VLBI images. If more than one image was available for a given target, a median $\mathrm{P.A._{jet}}$ was used. 
For 90\% of cases we estimated jet direction with accuracy better than $10^\circ$.
Variable direction of ejections \citep{lister_etal13}
or apparent jet curvature, enhanced by the projection effect \citep[e.g.,][]{Agudo07}, results in a larger error for remaining objects.
We succeeded in determining the jet orientation for 2957 matched AGNs; approximately half of the total number of matches.
A significant fraction of images did not have high-enough dynamic range to allow a robust jet direction determination.

Examples of VLBI images for four AGN targets overlayed with the $\vec{VG}$ and $\vec{j}$ vectors are shown in Fig~\ref{f:maps}. 
Since the VLBI images do not contain information on their absolute positions, for illustration purposes, the VLBI-\Gaia offset vector $\vec{VG}$ is shown relative to the peak intensity pixel on the maps.
We checked whether or not the VLBI-\Gaia offset position angle $\mathrm{P.A._{VG}}$ or parsec-scale radio jet $\mathrm{P.A._{jet}}$ have preferred directions on the sky and found that their distributions are flat over $360^\circ$.

\begin{figure}[t]
\centering
\includegraphics[width=0.5\textwidth,angle=0,trim=0.3cm 0.5cm 0cm 0.2cm,clip]{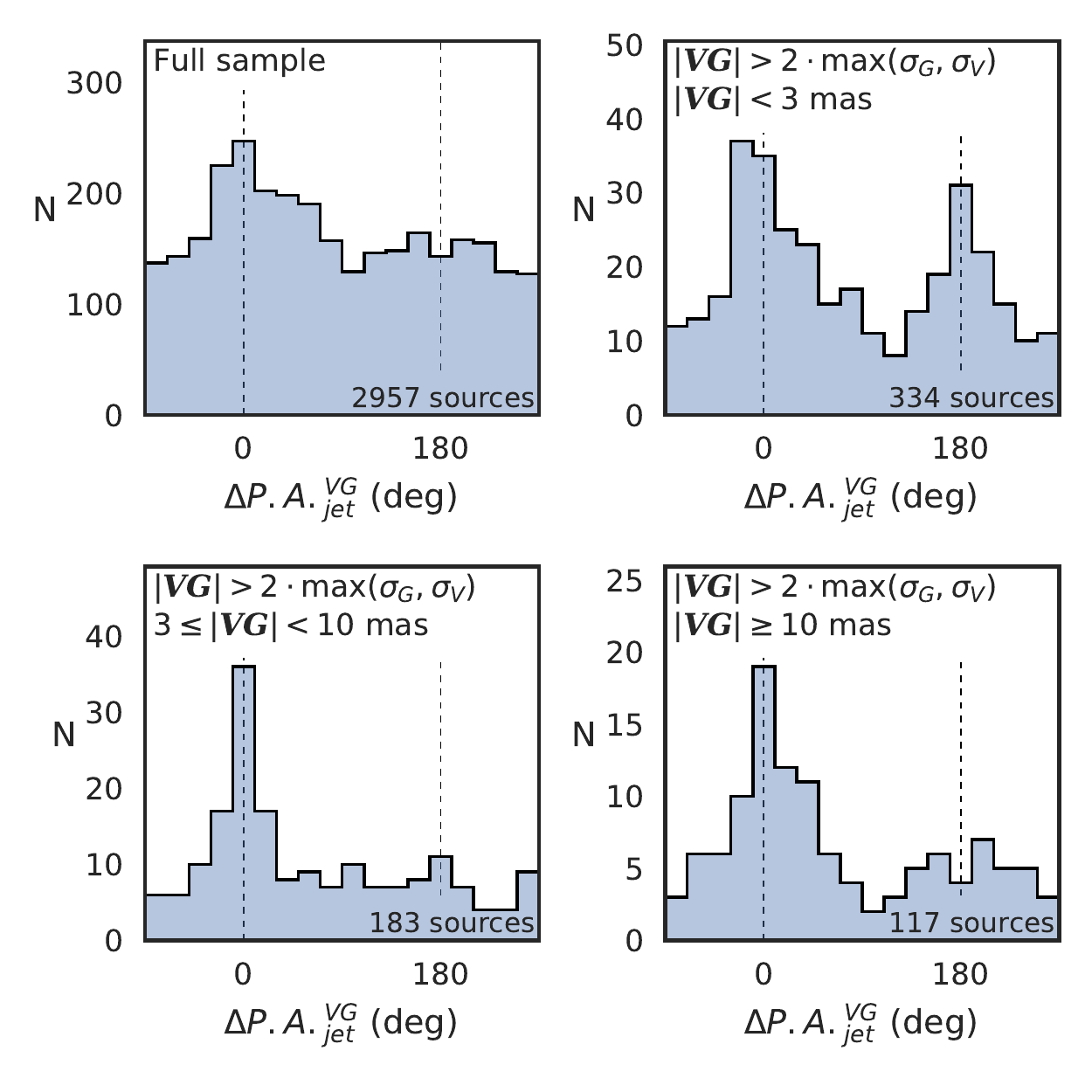}
\caption{
Distribution of $\Delta\mathrm{P.A.^{VG}_{jet}}$ for different sub-samples of VLBI-\Gaia associations with probability of false association less than $2\cdot10^{-4}$.
The vertical dashed lines are shown for $\Delta\mathrm{P.A.^{VG}_{jet}}=0^\circ$ (VLBI to \Gaia $\vec{VG}$ reference point offset vector along the jet) and $180^\circ$ ($\vec{VG}$ offset vector opposite to the jet direction) values.
\label{f:hist}
}
\end{figure}

\begin{figure}[t]
\centering
\includegraphics[width=0.5\textwidth,angle=0,trim=0.5cm 0cm 1cm 1cm,clip]{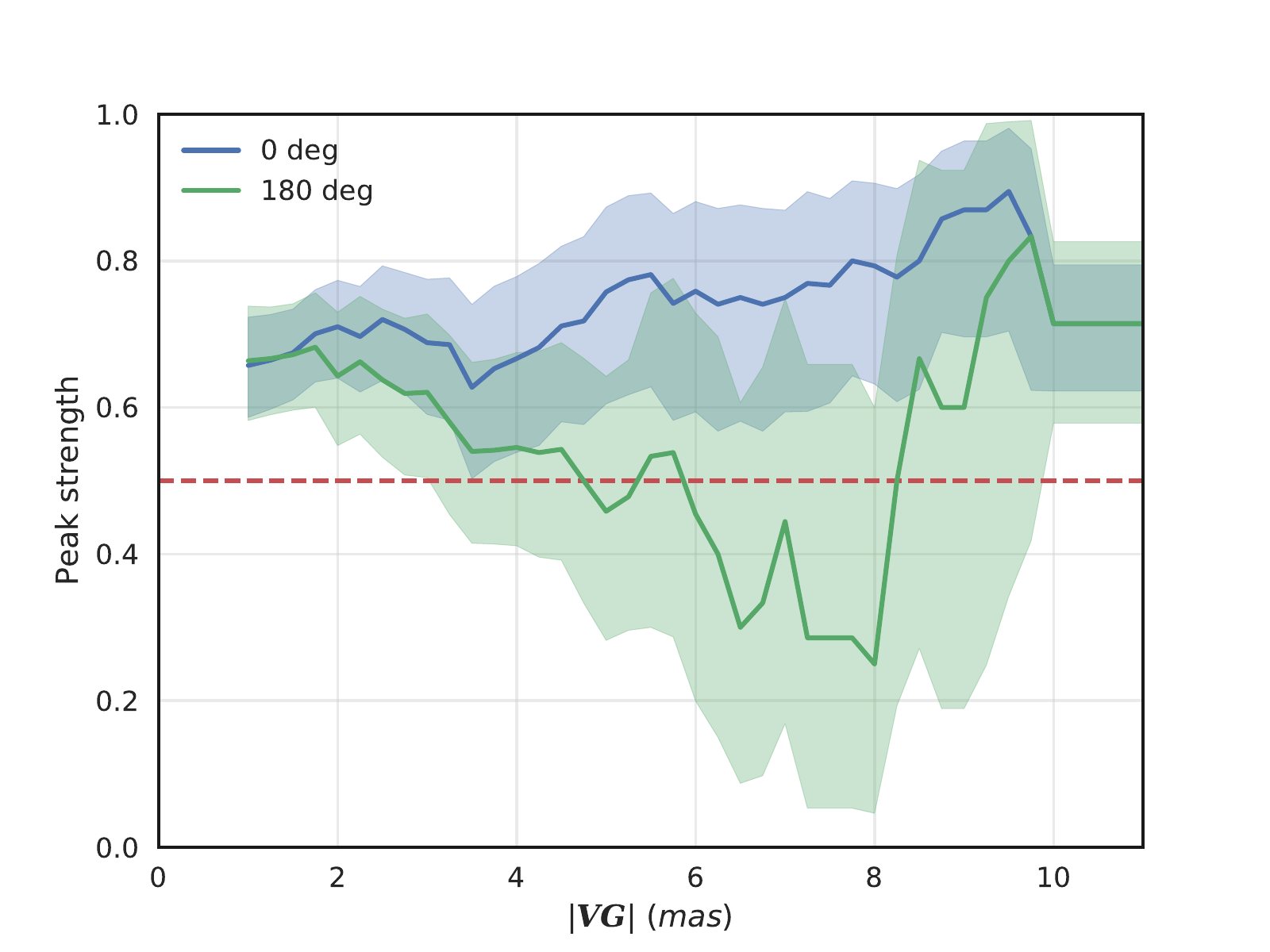}
\caption{
A measure of the peak strength in $\Delta\mathrm{P.A.^{VG}_{jet}}$ distributions calculated as discussed in \S\,\ref{s:direction}.
The horizontal dashed red line presents the flat distribution case.
The solid blue curve describes the $0^\circ$ peak, green color~-- the peak around  $180^\circ$.
Results for $|\vec{VG}|>10$~mas present an integration over all targets with $|\vec{VG}|>10$~mas. 
The shading shows the 90\,\% confidence interval.
\label{f:peaks}
}
\end{figure}

\section{Parsec-scale jet direction is preferred by the VLBI-\Gaia offset}
\label{s:direction}

The distribution of the differences $\Delta\mathrm{P.A.^{VG}_{jet}}$ between the VLBI-\Gaia offset position angle $\mathrm{P.A._{VG}}$ and the VLBI jet direction $\mathrm{P.A._{jet}}$ is presented in Fig.~\ref{f:hist}.
It is evident that $\Delta\mathrm{P.A.^{VG}_{jet}}$ prefers values $0^\circ$ and $180^\circ$ even for the full sample (top left histogram) with significance  $<10^{-8}$ and 0.007, respectively.
To further demonstrate this effect, we plot the other histograms in Fig.~\ref{f:hist}, only  for those associations that have the offset $|\vec{GV}|$ at least two times larger than the errors of their VLBI, $\sigma_\mathrm{V}$, and \Gaia, $\sigma_\mathrm{G}$, positions. For this sub-sample, the excess of targets that prefer the VLBI jet direction is estimated as 34\,\% and 13\,\% along and opposite to the jet direction, respectively. 
The excess may rise with improving \Gaia position accuracy in future data releases.
The histograms cover different typical intervals of the offset lengths
$|\vec{VG}|$, and clearly demonstrate different effects that are present at 
different offset scales. 
To present this in even more details, Fig.~\ref{f:peaks} shows a measure of the peak significance for the peak at $\Delta\mathrm{P.A.^{VG}_{jet}}=0^\circ$ and $\Delta\mathrm{P.A.^{VG}_{jet}}=180^\circ$ for associations with $\mathrm{PFA}<2\cdot10^{-4}$, $\sigma_\mathrm{G}<0.5\,|\vec{VG}|$, and $\sigma_\mathrm{V}< 0.5\,|\vec{VG}|$.
The measure is calculated as the number of targets within $\pm45^\circ$ from the peak divided by the total number of targets within $\pm90^\circ$ of the corresponding peak for distributions with offset values $\pm1$~mas from a given length $|\vec{VG}|$.
Binomial proportion intervals are calculated and shown for the 90\,\% confidence level.

We performed a Monte Carlo simulation in order to provide a quantitaive measure of the systematic VLBI-Gaia offset in the presence of the random noise.
The simulation aims to reproduce the peak at
$\Delta\mathrm{P.A.^{VG}_{jet}}=180^\circ$ 
in the histogram of the sub-sample with offset $|\vec{VG}|<3$~mas, 
$\sigma_\mathrm{G}<0.5\,|\vec{VG}|$, and $\sigma_\mathrm{V}< 0.5\,|\vec{VG}|$ (Fig.~\ref{f:hist}, top right).
In our model, the \Gaia positions were subject to two factors: systematic 
shift $s$ with respect to VLBI position that obeys the Gaussian distribution with 
the standard deviation $\sigma,$ and  random noise. The vector of random noise 
has uniform angular distribution and length that obeys the 
power-law transformed Rayleighian distribution with parameters found 
by \cite{PK16}. We ran one million trials at a grid over $s$ and $\sigma$. 
The probability of getting or exceeding the peak at histogram at $\Delta\mathrm{P.A.^{VG}_{jet}}=180^\circ$ 
at $s=0, \sigma=0$, that is,\ when the offsets have isotropic distribution, 
is $3 \cdot 10^{-3}$. The probability exceeds the confidence level 0.05 
either when $s > 0.06$~mas in the opposite jet direction or when $\sigma > 0.26$~mas.

We ran a similar simulation to reproduce the peak at
$\Delta\mathrm{P.A.^{VG}_{jet}}=0^\circ$ 
for the sub-sample with offset lengths $|\vec{VG}|>8$~mas, 
$\sigma_\mathrm{G}<0.5\,|\vec{VG}|$, and $\sigma_\mathrm{V}<0.5\,|\vec{VG}|$ (Fig.~\ref{f:peaks}).
The probability of getting or exceeding the peak at $\Delta\mathrm{P.A.^{VG}_{jet}}=0^\circ$ for $s=0, \sigma=0$, is less than $10^{-6}$. The probability exceeds the confidence 
level 0.05 either when $s > 1.2$~mas in the jet direction or when $\sigma > 2.6$~mas. This means
that this peak cannot be explained by the contribution of the radio source
structure to the shift of the VLBI reference point with respect to the jet base 
(`$\mathrm{C_r}$'-`$\mathrm{P_V}$' distance at the diagram in Fig.~\ref{f:diagram}), 
which is more than one order of magnitude less and, moreover, is expected for $\Delta\mathrm{P.A.^{VG}_{jet}}=180^\circ$.

Results presented here are confirmed on a lower significance level if the ICRF2 catalog \citep{ICRF2} or the \Gaia DR1 quasar auxiliary solution \citep{Gaia_DR1_frame} are used.

\section{Discussion}
\label{s:discussion}

Offsets in the jet direction ($\Delta\mathrm{P.A.^{VG}_{jet}}\approx0^\circ$) are observed for many matches for a wide range of VLBI-\Gaia distances $|\vec{VG}|$: from less than 1~mas to more than 10~mas (Figures~\ref{f:hist},\,\ref{f:peaks}).
See J0005+3820 and J1800+7828 in Fig.~\ref{f:maps} for an example.
We stress that the histogram of $\Delta\mathrm{P.A.^{VG}_{jet}}$ contradicts the assertion that the radio-optical offsets can be caused by the presence of extended-frequency-dependent parsec-scale radio structure alone. We must assume the presence of extended optical structure at milliarcsecond scales in order to explain the distributions in Figures~\ref{f:hist} and\,\ref{f:peaks}. Thus, we find  direct massive observational evidence of the existence of elongated, milliarsecond-scale optical jet structures. We note that observational evidence of a direct relation between optical and radio jet properties at parsec scales was discussed previously by, for example, \cite{marscher_etal08,marscher_etal10}.
The optical parsec-scale synchrotron AGN jets should be observed as extended and bright enough for a significant fraction of targets shifting the centroid of optical positions by more than 1~mas.
We note that some optical jets possess bright features even at a $1\arcsec$ distance from the nucleus \citep[see, e.g., the jet in M87 in][]{BSM99,Prieto16} which results in the observed significant shift in the optical centroid position relative to the VLBI-compact radio emission.
No clear distinction between parsec-scale structure of sources showing different typical $|\vec{VG}|$ and $\Delta\mathrm{P.A.^{VG}_{jet}}$ values was found.
%
%
%
Net rotation of \Gaia reference frame with respect to VLBI for the found AGN matches changes by less than 0.05 mas when the 384 AGNs with significant offsets \citep{PK16} are excluded.

Offsets opposite to the jet direction ($\Delta\mathrm{P.A.^{VG}_{jet}}\approx180^\circ$) are measured  mainly at small VLBI-\Gaia length $|\vec{VG}|<3$~mas and are not seen at larger distances, that is, $3<|\vec{VG}|<10$~mas.
An example of this case is presented in Fig.~\ref{f:maps} by J1229+0203.
Three reasons would favor this scenario, namely: (i) the core-shift effect, (ii) the effect of resolved VLBI jet structure, or (iii) the potential effect of optical emission from the accretion disk.
With the current accuracy of \Gaia positions and the lack of information of the \Gaia positional variability, we are unable to distinguish between these three scenarios. 
For future releases of \Gaia positions one could limit the analysis to the most compact VLBI targets to partially mitigate the effect of the parsec-scale VLBI structure of AGN jets.
%
%
\citet{Kovalev_cs_2008,Sullivan09_coreshift,pushkarev_etal12,Sokol_cs2011,MV16} have demonstrated that the core-shift can be precisely measured. \citet{C02} has shown how radio source structure can be accurately taken into account using their images.
After calibrating the radio position for core-shift and radio structure, the contribution of emission from the accretion disk and optical jet will be seen much more clearly. We anticipate observing programs targeting massive core-shift measurements and processing radio astrometry surveys with applied source structure term in the future, though this will require reconstruction of high-quality images within every VLBI astrometry session.

A weaker peak in the distribution is observed for VLBI-\Gaia shifts along the jet direction ($\Delta\mathrm{P.A.^{VG}_{jet}}\approx180^\circ$) for $|\vec{VG}|>10$~mas (Figs.~\ref{f:hist},\,\ref{f:peaks}). This might result from the influence of bright radio features far away from the central engine.

\section{Summary}

We find that the VLBI-\Gaia positional offsets prefer to be parallel to parsec-scale radio jet direction.

The offset from VLBI to \Gaia positions happens along the jet
in a range from a fraction of a mas to more than 10~mas.
In this case, optical centroids are farther away from the central nucleus.
This can only be explained if elongated bright optical jets exist at parsec scales in many AGNs and significantly shift the reference point from their apparent optical base.

Position offsets opposite to the jet direction do not exceed 3~mas and are as common as shifts along the jet.
For the opposite offsets, radio reference points are farther away from the central nucleus.
This could be due to several factors, including the apparent core-shift effect due to the synchrotron self-absorption, or the extended VLBI structure of radio jets.

\begin{acknowledgements}
We thank A.~B.~Pushkarev, A.~P.~Lobanov, E.~Ros, A.V.~Moiseev, G.V.~Lipunova, and the anonymous referee for useful discussions and suggestions.
This project is supported by the Russian Science Foundation grant 16-12-10481.
We deeply thank the teams referred to in \S~\ref{s:data-anal} for making their fully calibrated VLBI FITS data publicly available. This research has made use of data from the MOJAVE database that is maintained by the MOJAVE team \citep{MOJAVE_V}.
This study makes use of 43~GHz VLBA data from the VLBA-BU Blazar Monitoring Program, funded by NASA through the Fermi Guest Investigator Program.
This research has made use of NASA's Astrophysics Data System.
\end{acknowledgements}

\bibliographystyle{aa}
\bibliography{yyk}

\begin{thebibliography}{44}
\expandafter\ifx\csname natexlab\endcsname\relax\def\natexlab#1{#1}\fi

\bibitem[{{Agudo} {et~al.}(2007){Agudo}, {Bach}, {Krichbaum}, {Marscher},
  {Gonidakis}, {Diamond}, {Perucho}, {Alef}, {Graham}, {Witzel}, {Zensus},
  {Bremer}, {Acosta-Pulido}, \& {Barrena}}]{Agudo07}
{Agudo}, I., {Bach}, U., {Krichbaum}, T.~P., {et~al.} 2007, \aap, 476, L17

\bibitem[{{Beasley} {et~al.}(2002){Beasley}, {Gordon}, {Peck}, {Petrov},
  {MacMillan}, {Fomalont}, \& {Ma}}]{vcs1}
{Beasley}, A.~J., {Gordon}, D., {Peck}, A.~B., {et~al.} 2002, \apjs, 141, 13

\bibitem[{{Biretta} {et~al.}(1999){Biretta}, {Sparks}, \& {Macchetto}}]{BSM99}
{Biretta}, J.~A., {Sparks}, W.~B., \& {Macchetto}, F. 1999, \apj, 520, 621

\bibitem[{{Charlot}(1990)}]{Charlot90}
{Charlot}, P. 1990, \aj, 99, 1309

\bibitem[{{Charlot}(2002)}]{C02}
{Charlot}, P. 2002, in International VLBI Service for Geodesy and Astrometry:
  General Meeting Proceedings, ed. N.~R. {Vandenberg} \& K.~D. {Baver}, 233

\bibitem[{{Charlot} {et~al.}(2010){Charlot}, {Boboltz}, {Fey}, {Fomalont},
  {Geldzahler}, {Gordon}, {Jacobs}, {Lanyi}, {Ma}, {Naudet}, {Romney},
  {Sovers}, \& {Zhang}}]{Charlot10}
{Charlot}, P., {Boboltz}, D.~A., {Fey}, A.~L., {et~al.} 2010, \aj, 139, 1713

\bibitem[{{Falomo} {et~al.}(2000){Falomo}, {Scarpa}, {Treves}, \&
  {Urry}}]{2000ApJ...542..731F}
{Falomo}, R., {Scarpa}, R., {Treves}, A., \& {Urry}, C.~M. 2000, \apj, 542, 731

\bibitem[{{Feissel} {et~al.}(2000){Feissel}, {Gontier}, \& {Eubanks}}]{r:fes00}
{Feissel}, M., {Gontier}, A.-M., \& {Eubanks}, T.~M. 2000, \aap, 359, 1201

\bibitem[{{Fey} {et~al.}(2015){Fey}, {Gordon}, {Jacobs}, {Ma}, {Gaume},
  {Arias}, {Bianco}, {Boboltz}, {B{\"o}ckmann}, {Bolotin}, {Charlot},
  {Collioud}, {Engelhardt}, {Gipson}, {Gontier}, {Heinkelmann}, {Kurdubov},
  {Lambert}, {Lytvyn}, {MacMillan}, {Malkin}, {Nothnagel}, {Ojha},
  {Skurikhina}, {Sokolova}, {Souchay}, {Sovers}, {Tesmer}, {Titov}, {Wang}, \&
  {Zharov}}]{ICRF2}
{Fey}, A.~L., {Gordon}, D., {Jacobs}, C.~S., {et~al.} 2015, \aj, 150, 58

\bibitem[{{Fomalont} {et~al.}(2003){Fomalont}, {Petrov}, {MacMillan}, {Gordon},
  \& {Ma}}]{vcs2}
{Fomalont}, E.~B., {Petrov}, L., {MacMillan}, D.~S., {Gordon}, D., \& {Ma}, C.
  2003, \aj, 126, 2562

\bibitem[{{Fromm} {et~al.}(2013){Fromm}, {Ros}, {Perucho}, {Savolainen},
  {Mimica}, {Kadler}, {Lobanov}, \& {Zensus}}]{fromm_etal13}
{Fromm}, C.~M., {Ros}, E., {Perucho}, M., {et~al.} 2013, \aap, 557, A105

\bibitem[{{Helmboldt} {et~al.}(2007){Helmboldt}, {Taylor}, {Tremblay},
  {Fassnacht}, {Walker}, {Myers}, {Sjouwerman}, {Pearson}, {Readhead},
  {Weintraub}, {Gehrels}, {Romani}, {Healey}, {Michelson}, {Blandford}, \&
  {Cotter}}]{Hel07}
{Helmboldt}, J.~F., {Taylor}, G.~B., {Tremblay}, S., {et~al.} 2007, \apj, 658,
  203

\bibitem[{{Jorstad} \& {Marscher}(2016)}]{BU16}
{Jorstad}, S. \& {Marscher}, A. 2016, Galaxies, 4, 47

\bibitem[{{Kovalev} {et~al.}(2008){Kovalev}, {Lobanov}, {Pushkarev}, \&
  {Zensus}}]{Kovalev_cs_2008}
{Kovalev}, Y.~Y., {Lobanov}, A.~P., {Pushkarev}, A.~B., \& {Zensus}, J.~A.
  2008, \aap, 483, 759

\bibitem[{{Kovalev} {et~al.}(2007){Kovalev}, {Petrov}, {Fomalont}, \&
  {Gordon}}]{vcs5}
{Kovalev}, Y.~Y., {Petrov}, L., {Fomalont}, E.~B., \& {Gordon}, D. 2007, \aj,
  133, 1236

\bibitem[{{Kutkin} {et~al.}(2014){Kutkin}, {Sokolovsky}, {Lisakov}, {Kovalev},
  {Savolainen}, {Voytsik}, {Lobanov}, {Aller}, {Aller}, {Lahteenmaki},
  {Tornikoski}, {Volvach}, \& {Volvach}}]{kutkin_etal14}
{Kutkin}, A.~M., {Sokolovsky}, K.~V., {Lisakov}, M.~M., {et~al.} 2014, \mnras,
  437, 3396

\bibitem[{{Lindegren} {et~al.}(2016){Lindegren}, {Lammers}, {Bastian},
  {Hern{\'a}ndez}, {Klioner}, {Hobbs}, {Bombrun}, {Michalik}, {Ramos-Lerate},
  {Butkevich}, {Comoretto}, {Joliet}, {Holl}, {Hutton}, {Parsons},
  {Steidelm{\"u}ller}, {Abbas}, {Altmann}, {Andrei}, {Anton}, {Bach},
  {Barache}, {Becciani}, {Berthier}, {Bianchi}, {Biermann}, {Bouquillon},
  {Bourda}, {Br{\"u}semeister}, {Bucciarelli}, {Busonero}, {Carlucci},
  {Casta{\~n}eda}, {Charlot}, {Clotet}, {Crosta}, {Davidson}, {de Felice},
  {Drimmel}, {Fabricius}, {Fienga}, {Figueras}, {Fraile}, {Gai}, {Garralda},
  {Geyer}, {Gonz{\'a}lez-Vidal}, {Guerra}, {Hambly}, {Hauser}, {Jordan},
  {Lattanzi}, {Lenhardt}, {Liao}, {L{\"o}ffler}, {McMillan}, {Mignard}, {Mora},
  {Morbidelli}, {Portell}, {Riva}, {Sarasso}, {Serraller}, {Siddiqui}, {Smart},
  {Spagna}, {Stampa}, {Steele}, {Taris}, {Torra}, {van Reeven}, {Vecchiato},
  {Zschocke}, {de Bruijne}, {Gracia}, {Raison}, {Lister}, {Marchant},
  {Messineo}, {Soffel}, {Osorio}, {de Torres}, \&
  {O'Mullane}}]{Gaia_DR1_astrometry}
{Lindegren}, L., {Lammers}, U., {Bastian}, U., {et~al.} 2016, \aap, 595, A4

\bibitem[{{Liodakis} {et~al.}(2016){Liodakis}, {Pavlidou}, \&
  {Angelakis}}]{LPA16}
{Liodakis}, I., {Pavlidou}, V., \& {Angelakis}, E. 2016, \mnras, accepted;
  ArXiv e-prints [\eprint[arXiv]{1610.06561}]

\bibitem[{{Lister} {et~al.}(2009){Lister}, {Aller}, {Aller}, {Cohen}, {Homan},
  {Kadler}, {Kellermann}, {Kovalev}, {Ros}, {Savolainen}, {Zensus}, \&
  {Vermeulen}}]{MOJAVE_V}
{Lister}, M.~L., {Aller}, H.~D., {Aller}, M.~F., {et~al.} 2009, \aj, 137, 3718

\bibitem[{{Lister} {et~al.}(2013){Lister}, {Aller}, {Aller}, {Homan},
  {Kellermann}, {Kovalev}, {Pushkarev}, {Richards}, {Ros}, \&
  {Savolainen}}]{lister_etal13}
{Lister}, M.~L., {Aller}, M.~F., {Aller}, H.~D., {et~al.} 2013, \aj, 146, 120

\bibitem[{{Lobanov}(1998)}]{L98}
{Lobanov}, A.~P. 1998, \aap, 330, 79

\bibitem[{{Marcaide} \& {Shapiro}(1984)}]{MS84}
{Marcaide}, J.~M. \& {Shapiro}, I.~I. 1984, \apj, 276, 56

\bibitem[{{Marscher} {et~al.}(2008){Marscher}, {Jorstad}, {D'Arcangelo},
  {Smith}, {Williams}, {Larionov}, {Oh}, {Olmstead}, {Aller}, {Aller},
  {McHardy}, {L{\"a}hteenm{\"a}ki}, {Tornikoski}, {Valtaoja}, {Hagen-Thorn},
  {Kopatskaya}, {Gear}, {Tosti}, {Kurtanidze}, {Nikolashvili}, {Sigua},
  {Miller}, \& {Ryle}}]{marscher_etal08}
{Marscher}, A.~P., {Jorstad}, S.~G., {D'Arcangelo}, F.~D., {et~al.} 2008, \nat,
  452, 966

\bibitem[{{Marscher} {et~al.}(2010){Marscher}, {Jorstad}, {Larionov}, {Aller},
  {Aller}, {L{\"a}hteenm{\"a}ki}, {Agudo}, {Smith}, {Gurwell}, {Hagen-Thorn},
  {Konstantinova}, {Larionova}, {Larionova}, {Melnichuk}, {Blinov},
  {Kopatskaya}, {Troitsky}, {Tornikoski}, {Hovatta}, {Schmidt}, {D'Arcangelo},
  {Bhattarai}, {Taylor}, {Olmstead}, {Manne-Nicholas}, {Roca-Sogorb},
  {G{\'o}mez}, {McHardy}, {Kurtanidze}, {Nikolashvili}, {Kimeridze}, \&
  {Sigua}}]{marscher_etal10}
{Marscher}, A.~P., {Jorstad}, S.~G., {Larionov}, V.~M., {et~al.} 2010, \apjl,
  710, L126

\bibitem[{{Mart{\'{\i}}-Vidal} {et~al.}(2016){Mart{\'{\i}}-Vidal},
  {Abell{\'a}n}, {Marcaide}, {Guirado}, {P{\'e}rez-Torres}, \& {Ros}}]{MV16}
{Mart{\'{\i}}-Vidal}, I., {Abell{\'a}n}, F.~J., {Marcaide}, J.~M., {et~al.}
  2016, \aap, 596, A27

\bibitem[{{Mignard} {et~al.}(2016){Mignard}, {Klioner}, {Lindegren}, {Bastian},
  {Bombrun}, {Hern{\'a}ndez}, {Hobbs}, {Lammers}, {Michalik}, {Ramos-Lerate},
  {Biermann}, {Butkevich}, {Comoretto}, {Joliet}, {Holl}, {Hutton}, {Parsons},
  {Steidelm{\"u}ller}, {Andrei}, {Bourda}, \& {Charlot}}]{Gaia_DR1_frame}
{Mignard}, F., {Klioner}, S., {Lindegren}, L., {et~al.} 2016, \aap, 595, A5

\bibitem[{{O'Sullivan} \& {Gabuzda}(2009)}]{Sullivan09_coreshift}
{O'Sullivan}, S.~P. \& {Gabuzda}, D.~C. 2009, \mnras, 400, 26

\bibitem[{{Petrov}(2016)}]{VCS9}
{Petrov}, L. 2016, ArXiv e-prints [\eprint[arXiv]{1610.04951}]

\bibitem[{{Petrov} {et~al.}(2009){Petrov}, {Gordon}, {Gipson}, {MacMillan},
  {Ma}, {Fomalont}, {Walker}, \& {Carabajal}}]{RDV2009}
{Petrov}, L., {Gordon}, D., {Gipson}, J., {et~al.} 2009, Journal of Geodesy, 8

\bibitem[{{Petrov} \& {Kovalev}(2016)}]{PK16}
{Petrov}, L. \& {Kovalev}, Y.~Y. 2016, MNRAS submitted
  [\eprint[arXiv]{1611.02630}]

\bibitem[{{Petrov} {et~al.}(2005){Petrov}, {Kovalev}, {Fomalont}, \&
  {Gordon}}]{vcs3}
{Petrov}, L., {Kovalev}, Y.~Y., {Fomalont}, E.~B., \& {Gordon}, D. 2005, \aj,
  129, 1163

\bibitem[{{Petrov} {et~al.}(2006){Petrov}, {Kovalev}, {Fomalont}, \&
  {Gordon}}]{vcs4}
{Petrov}, L., {Kovalev}, Y.~Y., {Fomalont}, E.~B., \& {Gordon}, D. 2006, \aj,
  131, 1872

\bibitem[{{Petrov} {et~al.}(2008){Petrov}, {Kovalev}, {Fomalont}, \&
  {Gordon}}]{vcs6}
{Petrov}, L., {Kovalev}, Y.~Y., {Fomalont}, E.~B., \& {Gordon}, D. 2008, \aj,
  136, 580

\bibitem[{{Petrov} \& {Taylor}(2011)}]{Petrov_VIPS}
{Petrov}, L. \& {Taylor}, G.~B. 2011, AJ, 142, 89

\bibitem[{{Piner} {et~al.}(2012){Piner}, {Pushkarev}, {Kovalev}, {Marvin},
  {Arenson}, {Charlot}, {Fey}, {Collioud}, \& {Voitsik}}]{Piner12}
{Piner}, B.~G., {Pushkarev}, A.~B., {Kovalev}, Y.~Y., {et~al.} 2012, ApJ, 758,
  84

\bibitem[{{Porcas}(2009)}]{Porcas_cs2009}
{Porcas}, R.~W. 2009, \aap, 505, L1

\bibitem[{{Prieto} {et~al.}(2016){Prieto}, {Fern{\'a}ndez-Ontiveros},
  {Markoff}, {Espada}, \& {Gonz{\'a}lez-Mart{\'{\i}}n}}]{Prieto16}
{Prieto}, M.~A., {Fern{\'a}ndez-Ontiveros}, J.~A., {Markoff}, S., {Espada}, D.,
  \& {Gonz{\'a}lez-Mart{\'{\i}}n}, O. 2016, \mnras, 457, 3801

\bibitem[{{Pushkarev} {et~al.}(2012){Pushkarev}, {Hovatta}, {Kovalev},
  {Lister}, {Lobanov}, {Savolainen}, \& {Zensus}}]{pushkarev_etal12}
{Pushkarev}, A.~B., {Hovatta}, T., {Kovalev}, Y.~Y., {et~al.} 2012, \aap, 545,
  A113

\bibitem[{{Pushkarev} \& {Kovalev}(2012)}]{puskov12}
{Pushkarev}, A.~B. \& {Kovalev}, Y.~Y. 2012, \aap, 544, A34

\bibitem[{{Pushkarev} \& {Kovalev}(2015)}]{puskov15}
{Pushkarev}, A.~B. \& {Kovalev}, Y.~Y. 2015, \mnras, 452, 4274

\bibitem[{{Pushkarev} {et~al.}(2013){Pushkarev}, {Kovalev}, {Lister},
  {Hovatta}, {Savolainen}, {Aller}, {Aller}, {Ros}, {Zensus}, {Richards},
  {Max-Moerbeck}, \& {Readhead}}]{pushkarev_etal13}
{Pushkarev}, A.~B., {Kovalev}, Y.~Y., {Lister}, M.~L., {et~al.} 2013, \aap,
  555, A80

\bibitem[{{Schinzel} {et~al.}(2015){Schinzel}, {Petrov}, {Taylor}, {Mahony},
  {Edwards}, \& {Kovalev}}]{schinzel15}
{Schinzel}, F.~K., {Petrov}, L., {Taylor}, G.~B., {et~al.} 2015, \apjs, 217, 4

\bibitem[{{Sokolovsky} {et~al.}(2011){Sokolovsky}, {Kovalev}, {Pushkarev}, \&
  {Lobanov}}]{Sokol_cs2011}
{Sokolovsky}, K.~V., {Kovalev}, Y.~Y., {Pushkarev}, A.~B., \& {Lobanov}, A.~P.
  2011, \aap, 532, A38

\bibitem[{{Sovers} {et~al.}(2002){Sovers}, {Charlot}, {Fey}, \& {Gordon}}]{S02}
{Sovers}, O.~J., {Charlot}, P., {Fey}, A.~L., \& {Gordon}, D. 2002, in
  International VLBI Service for Geodesy and Astrometry: General Meeting
  Proceedings, ed. N.~R. {Vandenberg} \& K.~D. {Baver}, 243

\end{thebibliography}

\end{document}